# Universal Inverse Power law distribution for Fractal Fluctuations in Dynamical Systems: Applications for Predictability of Inter - annual Variability of Indian and USA Region Rainfall

A.M.Selvam

Deputy Director (Retired)
Indian Institute of Tropical Meteorology, Pune 411 008, India
Email: amselvam@gmail.com
Websites: http://amselvam.webs.com
http://amselvam.tripod.com/index.html

**Abstract**

Dynamical systems in nature exhibit self-similar fractal space-time fluctuations on all scales indicating long-range correlations and therefore the statistical normal distribution with implicit assumption of independence, fixed mean and standard deviation cannot be used for description and quantification of fractal data sets. The author has developed a general systems theory based on classical statistical physics for fractal fluctuations which predicts the following. (i) The fractal fluctuations signify an underlying eddy continuum, the larger eddies being the integrated mean of enclosed smaller-scale fluctuations. (ii) The probability distribution of eddy amplitudes and the variance (square of eddy amplitude) spectrum of fractal fluctuations follow the universal Boltzmann inverse power law expressed as a function of the golden mean. (iii) Fractal fluctuations are signatures of quantum-like chaos since the additive amplitudes of eddies when squared represent probability densities analogous to the sub-atomic dynamics of quantum systems such as the photon or electron. (iv) The model predicted distribution is very close to statistical normal distribution for moderate events within two standard deviations from the mean but exhibits a fat long tail that are associated with hazardous extreme events. Continuous periodogram power spectral analyses of available GHCN annual total rainfall time series for the period 1900 to 2008 for Indian and USA stations show that the power spectra and the corresponding probability distributions follow model predicted universal inverse power law form signifying an eddy continuum structure underlying the observed inter-annual variability of rainfall. Global warming related atmospheric energy input will result in intensification of fluctuations of all scales and can be seen immediately in high frequency (short-term) fluctuations such as devastating floods/droughts resulting from excess/deficit annual, quasi-biennial and other shorter period (years) rainfall cycles.

## 1. Introduction

Atmospheric flows exhibit self-similar fractal fluctuations on all space-time scales ranging from turbulence scale of a few millimeters-seconds to planetary scale of thousands of kilometers-years. Fractal space-time fluctuations are ubiquitous to dynamical systems in nature such as fluid flows, population growth, stock market indices, heart beat patterns, etc. (Mandelbrot, 1975). The power (variance) spectra of fractal fluctuations follow inverse power law, also called 1/$f$ noise, in the form $f^{\alpha}$ where $f$ is the frequency and α the exponent and imply long-range space-time correlations since the variance (intensity of fluctuations) is a function of frequency $f$ alone for the frequency range for which α is a constant. The study of power laws spans many disciplines, including physics, biology, engineering, computer science, the earth sciences, economics, political science, sociology, and statistics (Clauset et al., 2009; Kaniadakis, 2009). The observed scale invariance or long-range space-time correlations imply inherent 'persistence' or 'memory' in the space-time fluctuation patterns and are identified as signatures of self-organized criticality (Bak et al., 1988) intrinsic to dynamical systems in nature.

Lovejoy and Schertzer (2010) have given an exhaustive account of the observed scale invariant characteristics of atmospheric flows and emphasize the urgent need to incorporate the observed inverse power law scaling concepts in atmospheric sciences as summarized in the following. In spite of the unprecedented quantity and quality of meteorological data and numerical models, there is still no consensus about the atmosphere's elementary statistical properties as functions of scale in either time or in space. At present, the null hypotheses are classical so that they assume there are no long range statistical dependencies and that the probabilities are thin-tailed (i.e. exponential). However we have seen that cascades involve long range dependencies and (typically) have fat tailed (algebraic) distributions in which extreme events occur much more frequently and can persist for much longer than classical theory would allow.

The question of which statistical model best describes internal climate variability on interannual and longer time scales is essential to the ability to predict such variables and detect periodicities and trends in them. For over 30 years the dominant model for background climate variability has been the autoregressive model of the first order (AR1). However, recent research has shown that some aspects of climate variability are best described by a "long memory" or "power-law" model. Such a model fits a temporal spectrum to a single power-law function, which thereby accumulates more power at lower frequencies than an AR1 fit. Power-law behavior has been observed in globally and hemispherically averaged surface air temperature (Bloomfield 1992; Gil-Alana 2005), station surface air temperature (Pelletier 1997), geopotential height at 500 hPa (Tsonis et al. 1999), temperature paleoclimate proxies (Pelletier 1997; Huybers and Curry 2006), and many other studies (Vyushin and Kushner, 2009).

A general systems theory originally developed for atmospheric flows by Selvam (1990, 2005, 2007, 2009, 2010) predicts the observed self-organized criticality as a direct consequence of quantum-like chaos exhibited by fractal fluctuations generic to dynamical systems in nature. The model further predicts that the distribution of fractal fluctuations and the power spectrum (of fractal fluctuations) follow the same inverse power law which is a function of the golden mean $\tau$ ($\approx 1.618$). Model predictions are in agreement with continuous periodogram power spectral analyses of annual rainfall time series for Indian and USA region stations obtained from The Global Historical Climatology Network (GHCN-Monthly) of the National Oceanic and Atmospheric Administration's National Climate Data Center data base for the period 1900 to 2008. The paper is organized as follows. The general systems theory for self-similar fractal fluctuations is summarized in Section 2 and the application of classical statistical physics principles in general systems theory for the derivation of Boltzmann probability distribution for fractal fluctuations is discussed in Section 3. Details of data sets used for the study are given in Section 4. Analysis techniques and results are described in Section 5. Discussions of results and conclusions from the study are presented in Section 6.

## 2. General systems theory for fractal fluctuations in dynamical systems

The inverse power law form for power spectra of fractal fluctuations signifies an eddy continuum underlying the apparently irregular (unpredictable) fluctuation pattern. The fractal fluctuations may be visualized to result from the superimposition of a continuum of eddies (waves), the larger eddies enclosing the smaller eddies, i.e., the space-time integration of enclosed smaller eddies gives rise to formation of successively larger eddies. Such a simple concept of generation of large scale fluctuations from the integrated mean of inherent ubiquitous small-scale (turbulent) fluctuations gives the following equation (Townsend, 1956) for the relationship between the eddy circulation speeds $W$ and $w_*$ of large and turbulent eddies respectively and their corresponding radii $R$ and $r$.

$$W^2 = \frac{2}{\pi}\frac{r}{R}w_*^2 \qquad (1)$$

The above Eq. (1) represents the basic concepts underlying the general systems theory and is the governing equation for the growth of successively larger scale eddies resulting from the integrated mean of enclosed smaller scale eddies leading to the formation of an eddy continuum as explained in the following. The growth of the large eddy results from the cooperative existence of internal small scale eddies. Since the square of eddy circulation speeds $W^2$ and $w_*^2$ represent eddy energies (kinetic), the above equation also quantifies the ordered two-way eddy energy flow between the larger and smaller scales in terms of the length scale ratio $z$ equal to $r/R$ and is independent of any other physical, chemical, electrical properties of the medium of propagation. Large eddy growth exhibits the complex dynamics of a fuzzy logic system which responds as a unified whole to a multitude of inputs. The signatures of internal smaller scale fluctuations are carried as fine scale structure of large eddy circulations and contribute to the long-term correlations or 'memory' exhibited by dynamical systems.

## 2.1 The probability density distribution of fractal fluctuations

Statistical and mathematical tools are used for analysis of data sets and estimation of the probabilities of occurrence of events of different magnitudes in all branches of science and other areas of human interest. Historically, the statistical normal or the Gaussian distribution has been in use for nearly 400 years and gives a good estimate for probability of occurrence of the more frequent moderate sized events of magnitudes within two standard deviations from the mean. The Gaussian distribution is based on the concept of data independence, fixed mean and standard deviation with a majority of data events clustering around the mean. However, for real world infrequent hazardous extreme events of magnitudes greater than two standard deviations, the statistical normal distribution gives progressively increasing under-estimates of up to near zero probability. In the 1890s the power law or Pareto distributions with implicit long-range correlations were found to fit the fat tails exhibited by hazardous extreme events such as heavy rainfall, stock market crashes, traffic jams, the after-shocks following major earthquakes, etc. A historical review of statistical normal and the Pareto distributions are given by Andriani and McKelvey (2007) and Selvam (2009). The spatial and/or temporal data sets in practice refer to real world or computed dynamical systems and are fractals with self-similar geometry and long-range correlations in space and/or time, i.e., the statistical properties such as the mean and variance are scale-dependent and do not possess fixed mean and variance and therefore the statistical normal distribution cannot be used to quantify/describe self-similar data sets. Though the observed power law distributions exhibit qualitative universal shape, the exact physical mechanism underlying such scale-free power laws is not yet identified for the formulation of universal quantitative equations for fractal fluctuations of all scales. In the following Sec. 2.2 the universal inverse power law for fractal fluctuations is shown to be a function of the golden mean based on general systems theory concepts for fractal fluctuations.

## 2.2 Model predictions

The general systems theory model predictions for the space-time fractal fluctuation pattern of dynamical systems (Selvam, 1900, 2005, 2007, 2009, 2010) are given in the following

### 2.2.1 Quasiperiodic Penrose tiling pattern underlying fractal fluctuations

The power spectra of fractal fluctuations follow inverse power law form signifying an underlying eddy continuum structure. Visualization of large eddies as envelopes enclosing internal small scale eddies leads to the result that the successive eddy length/time scales of component eddies and their circulation speeds in the eddy continuum follow the Fibonacci mathematical number series such that the ratio of successive eddy radii $R_{n+1}/R_n$ and also circulation speeds $W_{n+1}/W_n$ is equal to the golden mean $\tau$ ($\approx 1.618$).

The apparently irregular fractal fluctuations can be resolved into a precise geometrical pattern with logarithmic spiral trajectory and the quasi periodic Penrose tiling pattern (Steinhardt, 1997) for the internal structure (Fig. 1) on all scales to form a nested continuum of vortex roll circulations with ordered energy flow between the scales (Eq. 1). A hierarchy of logarithmic spiral circulations contributes to the formation of the observed self-similar fractal fluctuations in dynamical systems.

The spiral flow structure $OR_OR_1R_2R_3R_4R_5$ can be visualized as an eddy continuum generated by successive length step growths $OR_O$, $OR_1$, $OR_2$, $OR_3$,….respectively equal to $R_1$, $R_2$, $R_3$,….which follow *Fibonacci* mathematical series such that $R_{n+1}=R_n+R_{n-1}$ and $R_{n+1}/R_n=\tau$ where $\tau$ is the *golden mean* equal to $(1+\sqrt{5})/2$ ($\approx 1.618$). Considering a normalized length step equal to 1 for the last stage of eddy growth, the successively decreasing radial length steps can be expressed as 1, $1/\tau$, $1/\tau^2$, $1/\tau^3$, ……The normalized eddy continuum comprises of fluctuation length scales 1, $1/\tau$, $1/\tau^2$, …….. The probability of occurrence is equal to $1/\tau$ and $1/\tau^2$ respectively for eddy length scale $1/\tau$ in any one or both rotational (clockwise and anti-clockwise) directions. Eddy fluctuation length of amplitude $1/\tau$ has a probability of occurrence equal to $1/\tau^2$ in both rotational directions, i.e., the square of eddy amplitude represents the probability of occurrence in the eddy continuum. Similar result is observed in the subatomic dynamics of quantum systems which are visualized to consist of the superimposition of eddy fluctuations in wave trains (eddy continuum).

The eddy continuum underlying fractal fluctuations has embedded robust dominant wavebands $R_OOR_1$, $R_1R_2O$, $R_3R_2O$, $R_3R_4O$, …… with length (time) scales $T_D$ which are functions of the golden mean $\tau$ and the primary eddy energy perturbation length (time) scale $T_S$ such as the annual cycle of summer to winter solar heating in atmospheric flows. The dominant eddy length (time) scale for the $n^{th}$ dominant eddy is given as

$$T_D = T_S(2+\tau)\tau^n \qquad (2)$$

The successive dominant eddy length (time) scales for unit primary perturbation length (time) scale, i.e. $T_S = 1$, are given (Eq. 2) as 2.2, 3.6, 5.9, 9.5, 15.3, 24.8, 40.1, 64.9, …respectively for values of $n$ = -1, 0, 1, 2, 3, 4, 5, 6,….

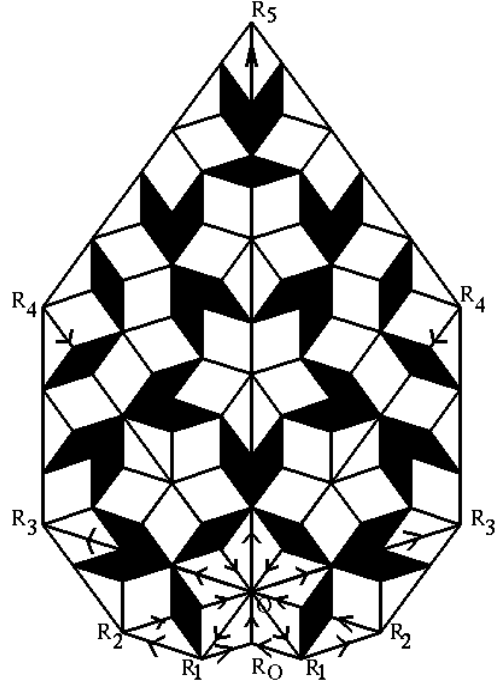

Fig. 1: The quasi-periodic Penrose tiling pattern traced by
the internal flow pattern of large eddy circulations

## 2.2.2 Logarithmic spiral pattern underlying fractal fluctuations

The overall logarithmic spiral flow structure $OR_OR_1R_2R_3R_4R_5$ (Fig. 1) is given by the relation

$$W = \frac{w_*}{k} \ln z \qquad (3)$$

In Eq. (3) the constant $k$ is the steady state fractional volume dilution of large eddy by inherent turbulent eddy fluctuations and $z$ is the length scale ratio $R/r$. The constant $k$ is equal to $1/\tau^2$ ($\cong 0.382$) and is identified as the universal constant for deterministic chaos in fluid flows. The steady state emergence of fractal structures is therefore equal to

$$\frac{1}{k} = \frac{WR}{w_* r} \cong 2.62 \qquad (4)$$

In Eq. (3), $W$ represents the standard deviation of eddy fluctuations, since $W$ is computed as the instantaneous r. m. s. (root mean square) eddy perturbation amplitude with reference to the earlier step of eddy growth. For two successive stages of eddy growth starting from primary perturbation $w_*$, the ratio of the standard deviations $W_{n+1}$ and $W_n$ is given from Eq. (3) as $(n+1)/n$. Denoting by $\sigma$ the standard deviation of eddy fluctuations at the reference level ($n=1$) the standard deviations of eddy fluctuations for successive stages of eddy growth are given as integer multiples of $\sigma$, i.e., $\sigma$, $2\sigma$, $3\sigma$, etc. and correspond respectively to

$$\text{statistical normalised standard deviation } t = 0, 1, 2, 3, .... \qquad (5)$$

The conventional power spectrum plotted as the variance versus the frequency in log-log scale will now represent the eddy probability density on logarithmic scale versus the standard deviation of the eddy fluctuations on linear scale since the logarithm of the eddy wavelength represents the standard deviation, i.e., the r. m. s. value of eddy fluctuations (Eq. 3). The r. m. s. value of eddy fluctuations can be represented in terms of statistical normal distribution as follows. A normalized standard deviation $t=0$ corresponds to cumulative percentage probability density equal to 50 for the mean value of the distribution. Since the logarithm of the wavelength represents the r. m. s. value of eddy fluctuations the normalized standard deviation $t$ is defined for the eddy energy as

$$t = \frac{\log L}{\log T_{50}} - 1 \tag{6}$$

In Eq. (6) $L$ is the time period (or wavelength) and $T_{50}$ is the period up to which the cumulative percentage contribution to total variance is equal to 50 and $t = 0$. $\log T_{50}$ also represents the mean value for the r. m. s. eddy fluctuations and is consistent with the concept of the mean level represented by r. m. s. eddy fluctuations. Spectra of time series of meteorological parameters when plotted as cumulative percentage contribution to total variance versus normalized deviation $t$ have been shown to follow closely the model predicted universal spectrum (Selvam and Fadnavis, 1998; Joshi and Selvam, 1999) which is identified as a signature of quantum-like chaos. The model predicted $T_{50}$ is obtained from Eq. (2) as equal to 3.6 years (Eq. 7) for the annual rainfall time series used in the present study where the primary perturbation time period $T_S$ is equal to one year (the annual cycle of summer to winter cycle of solar heating).

$$T_{50} = T_S(2+\tau)\tau^n = (2+\tau)\tau^0 \approx 3.6 \tag{7}$$

### 2.2.3 Universal Feigenbaum's constants and probability distribution function for fractal fluctuations

Selvam (1993, 2007) has shown that Eq. (1) represents the universal algorithm for deterministic chaos in dynamical systems and is expressed in terms of the universal *Feigenbaum's* (1980) *constants a* and *d* as follows. The successive length step growths generating the eddy continuum $OR_OR_1R_2R_3R_4R_5$ (Fig. 1) analogous to the period doubling route to chaos (growth) is initiated and sustained by the turbulent (fine scale) eddy acceleration $w_*$, which then propagates by the inherent property of inertia of the medium of propagation. Therefore, the statistical parameters *mean, variance, skewness* and *kurtosis* of the perturbation field in the medium of propagation are given by $w_*, w_*^2, w_*^3$ and $w_*^4$ respectively. The associated dynamics of the perturbation field can be described by the following parameters. The perturbation speed $w_*$ (motion) per second (unit time) sustained by its inertia represents the mass, $w_*^2$ the acceleration or force, $w_*^3$ the angular momentum or potential energy, and $w_*^4$ the spin angular momentum, since an eddy motion has an inherent curvature to its trajectory.

It is shown that *Feigenbaum's* constant $a$ is equal to (Selvam, 1993, 2007)

$$a = \frac{W_2 R_2}{W_1 R_1} \tag{8}$$

In Eq. (8) the subscripts 1 and 2 refer to two successive stages of eddy growth. *Feigenbaum's* constant $a$ as defined above represents the steady state emergence of fractional *Euclidean* structures. Considering dynamical eddy growth processes, *Feigenbaum's* constant $a$ also represents the steady state fractional outward mass dispersion rate and $a^2$ represents the energy flux into the environment generated by the persistent primary perturbation $W_1$. Considering both clockwise and counterclockwise rotations, the total energy flux into the environment is equal to $2a^2$. In statistical terminology, $2a^2$ represents the variance of fractal structures for both clockwise and counterclockwise rotation directions.

The probability of occurrence $P_{tot}$ of fractal domain $W_1R_1$ in the total larger eddy domain $W_nR_n$ in any (irrespective of positive or negative) direction is equal to

$$P_{tot} = \frac{W_1R_1}{W_nR_n} = \tau^{-2n}$$

Therefore the probability $P$ of occurrence of fractal domain $W_1R_1$ in the total larger eddy domain $W_nR_n$ in any one direction (either positive or negative) is equal to

$$P = \left(\frac{W_1R_1}{W_nR_n}\right)^2 = \tau^{-4n} \qquad (9)$$

The *Feigenbaum's* constant $d$ is shown to be equal to (Selvam, 1993, 2007)

$$d = \frac{W_2^4 R_2^3}{W_1^4 R_1^3} \qquad (10)$$

Eq. (10) represents the fractional volume intermittency of occurrence of fractal structures for each length step growth. *Feigenbaum's* constant $d$ also represents the relative spin angular momentum of the growing large eddy structures as explained earlier.

Eq. (1) may now be written as

$$2\frac{W^2 R^2}{w_*^2 (dR)^2} = \pi \frac{W^4 R^3}{w_*^4 (dR)^3} \qquad (11)$$

In Eq. (11) d$R$ equal to $r$ represents the incremental growth in radius for each length step growth, i.e., $r$ relates to the earlier stage of eddy growth.

The Feigenbaum's constant $d$ represented by $R/r$ is equal to

$$d = \frac{W^4 R^3}{w_*^4 r^3} \qquad (12)$$

For two successive stages of eddy growth

$$d = \frac{W_2^4 R_2^3}{W_1^4 R_1^3} \qquad (13)$$

From Eq. (1)

$$W_1^2 = \frac{2}{\pi}\frac{r}{R_1} w_*^2$$

$$W_2^2 = \frac{2}{\pi}\frac{r}{R_2} w_*^2 \qquad (14)$$

Therefore

$$\frac{W_2^2}{W_1^2} = \frac{R_1}{R_2} \qquad (15)$$

Substituting in Eq. (13)

$$d = \frac{W_2^4 R_2^3}{W_1^4 R_1^3} = \frac{W_2^2}{W_1^2} \frac{W_2^2 R_2^3}{W_1^2 R_1^3} = \frac{R_1}{R_2} \frac{W_2^2 R_2^3}{W_1^2 R_1^3} = \frac{W_2^2 R_2^2}{W_1^2 R_1^2} \qquad (16)$$

The Feigenbaum's constant $d$ represents the scale ratio $R_2/R_1$ and the inverse of the Feigenbaum's constant $d$ equal to $R_1/R_2$ represents the probability $(Prob)_1$ of occurrence of length scale $R_1$ in the total fluctuation length domain $R_2$ for the first eddy growth step as given in the following

$$(Prob)_1 = \frac{R_1}{R_2} = \frac{1}{d} = \frac{W_1^2 R_1^2}{W_2^2 R_2^2} = \tau^{-4} \qquad (17)$$

In general for the $n^{th}$ eddy growth step, the probability $(Prob)_n$ of occurrence of length scale $R_1$ in the total fluctuation length domain $R_n$ is given as

$$(Prob)_n = \frac{R_1}{R_n} = \frac{W_1^2 R_1^2}{W_n^2 R_n^2} = \tau^{-4n} \qquad (18)$$

The above equation for probability $(Prob)_n$ also represents, for the $n^{th}$ eddy growth step, the following statistical and dynamical quantities of the growing large eddy with respect to the initial perturbation domain: (i) the statistical relative variance of fractal structures, (ii) probability of occurrence of fractal domain in either positive or negative direction, and (iii) the inverse of $(Prob)_n$ represents the organized fractal (fine scale) energy flux in the overall large scale eddy domain. Large scale energy flux therefore occurs not in bulk, but in organized internal fine scale circulation structures identified as fractals.

Substituting the *Feigenbaum's constants* $a$ and $d$ defined above (Eqs. 8 and 10), Eq. (11) can be written as

$$2a^2 = \pi d \qquad (19)$$

In Eq. (19) $\pi d$, the relative volume intermittency of occurrence contributes to the total variance $2a^2$ of fractal structures.

In terms of eddy dynamics, the above equation states that during each length step growth, the energy flux into the environment equal to $2a^2$ contributes to generate relative spin angular momentum equal to $\pi d$ of the growing fractal structures. Each length step growth is therefore associated with a factor of $2a^2$ equal to $2\tau^4$ ($\cong 13.70820393$) increase in energy flux in the associated fractal domain. Ten such length step growths results in the formation of robust (self-sustaining) dominant bidirectional large eddy circulation $OR_OR_1R_2R_3R_4R_5$ (Fig. 1) associated with a factor of $20a^2$ equal to $137.08203$ increase in eddy energy flux. This non-dimensional constant factor characterizing successive dominant eddy energy increments is analogous to the *fine structure* constant $\propto^{-1}$ (Ford, 1968) observed in atomic spectra, where the spacing (energy) intervals between adjacent spectral lines is proportional to the non-dimensional *fine structure* constant equal to approximately 1/137. Further, the probability of $n^{th}$ length step eddy growth is given by $a^{-2n}$ ($\cong 6.8541^{-n}$) while the associated increase in eddy energy flux into the environment is equal to $a^{2n}$ ($\cong 6.8541^n$). Extreme events occur for large number of length step growths $n$ with small probability of occurrence and are associated with large energy release in the fractal domain. Each length step growth is associated with one-tenth of *fine structure constant* energy increment equal to $2a^2$ ($\alpha^{-1}/10 \cong 13.7082$) for bidirectional eddy circulation, or equal to one-twentieth of *fine structure constant* energy increment equal to $a^2$ ($\alpha^{-1}/20 \cong 6.8541$) in any one direction, i.e., positive or negative. The energy increase between two successive eddy length step growths may be expressed as a function of $(a^2)^2$, i.e., proportional to the square of the *fine structure constant* $\alpha^{-1}$. In the spectra of many atoms, what appears with coarse observations to be a single spectral line proves, with finer observation, to be a group of two or more closely spaced lines. The spacing of these fine-structure lines relative to the coarse spacing in the spectrum is proportional to the square of *fine structure constant*, for which reason this combination is called the *fine-structure constant*. We now know that the significance of the *fine-structure constant* goes beyond atomic spectra (Ford, 1968).

It was shown at Eq. (4) (Sec. 2.2.2) above that the steady state emergence of fractal structures in fluid flows is equal to $1/k$ ($=\tau^2$) and therefore the *Feigenbaum's constant a* is equal to

$$a = \tau^2 = \frac{1}{k} = 2.62 \qquad (20)$$

**2.2.4 Universal Feigenbaum's constants and power spectra of fractal fluctuations**

The power spectra of fluctuations in fluid flows can now be quantified in terms of universal *Feigenbaum's constant a* as follows.

The normalized variance and therefore the statistical probability distribution is represented by (from Eq. 9)

$$P = a^{-2t} \qquad (21)$$

In Eq. (21) $P$ is the probability density corresponding to normalized standard deviation $t$. The graph of $P$ versus $t$ will represent the power spectrum. The slope $Sl$ of the power spectrum is equal to

$$Sl = \frac{dP}{dt} \approx -P \qquad (22)$$

The power spectrum therefore follows inverse power law form, the slope decreasing with increase in $t$. Increase in $t$ corresponds to large eddies (low frequencies) and is consistent with observed decrease in slope at low frequencies in dynamical systems.

The probability distribution of fractal fluctuations (Eq. 18) is therefore the same as variance spectrum (Eq. 21) of fractal fluctuations.

The steady state emergence of fractal structures for each length step growth for any one direction of rotation (either clockwise or anticlockwise) is equal to

$$\frac{a}{2} = \frac{\tau^2}{2}$$

since the corresponding value for both direction is equal to $a$ (Eqs. 4 and 20 ).

The emerging fractal space-time structures have moment coefficient of kurtosis given by the fourth moment equal to

$$\left(\frac{\tau^2}{2}\right)^4 = \frac{\tau^8}{16} = 2.9356 \approx 3$$

The moment coefficient of skewness for the fractal space-time structures is equal to zero for the symmetric eddy circulations. Moment coefficient of kurtosis equal to 3 and moment coefficient of skewness equal to *zero* characterize the statistical normal distribution. The model predicted power law distribution for fractal fluctuations is close to the Gaussian distribution.

### 2.2.5 The power spectrum and probability distribution of fractal fluctuations are the same

The relationship between *Feigenbaum's constant a* and power spectra may also be derived as follows.

The steady state emergence of fractal structures is equal to the *Feigenbaum's constant a* (Eqs. 4 and 20). The relative variance of fractal structure which also represents the probability *P* of occurrence of bidirectional fractal domain for each length step growth is then equal to $1/a^2$. The normalized variance $\frac{1}{a^{2n}}$ will now represent the statistical probability density for the $n^{th}$ step growth according to model predicted quantum-like chaos for fluid flows. Model predicted probability density values *P* are computed as

$$P = \frac{1}{a^{2n}} = \tau^{-4n} \qquad (23)$$

or

$$P = \tau^{-4t} \qquad (24)$$

In Eq. (24) *t* is the normalized standard deviation (Eq. 5) for values of $t \geq 1$ and $t \leq -1$. The model predicted *P* values corresponding to normalized deviation *t* values less than 2 are slightly less than the corresponding statistical normal distribution values while the *P* values are noticeably larger for normalized deviation *t* values greater than 2 (Table 1 and Fig. 2) and may explain the reported *fat tail* for probability distributions of various physical parameters (Buchanan, 2004).

Values of the normalized deviation *t* in the range $-1 < t < 1$ refer to regions of primary eddy growth where the fractional volume dilution *k* (Eq. 4) by eddy mixing process has to be taken into account for determining the probability distribution *P* of fractal fluctuations (see Sec. 2.2.6 below).

### 2.2.6 Primary eddy growth region fractal space-time fluctuation probability distribution

Normalized deviation *t* ranging from -1 to +1 corresponds to the primary eddy growth region. In this region the probability *P* is shown to be equal to $P = \tau^{-4k}$ (see below) where *k* is the fractional volume dilution by eddy mixing (Eq. 4).

The normalized deviation *t* represents the length step growth number for growth stage more than one. The first stage of eddy growth is the primary eddy growth starting from unit length scale ($r = 1$) perturbation, the complete eddy forming at the tenth length scale growth, i.e., $R = 10r$ and scale ratio *z* equals 10 (Selvam, 1990). The steady state fractional volume

dilution *k* of the growing primary eddy by internal smaller scale eddy mixing is given from Eq. (4) as

$$k = \frac{w_* r}{WR} \qquad (25)$$

The expression for *k* in terms of the length scale ratio *z* equal to *R/r* is obtained from Eq. (1) as

$$k = \sqrt{\frac{\pi}{2z}} \qquad (26)$$

A fully formed primary large eddy length $R = 10r$ ($z=10$) represents the average or mean level zero and corresponds to a maximum of 50% probability of occurrence of either positive or negative fluctuation peak at normalized deviation *t* value equal to zero by convention. For intermediate eddy growth stages, i.e., *z* less than 10, the probability of occurrence of the primary eddy fluctuation does not follow conventional statistics, but is computed as follows taking into consideration the fractional volume dilution of the primary eddy by internal turbulent eddy fluctuations. Starting from unit length scale fluctuation, the large eddy formation is completed after 10 unit length step growths, i.e., a total of 11 length steps including the initial unit ($r = 1$) perturbation. At the second step ($z = R/r = 2$) of eddy growth the value of normalized deviation *t* is equal to 1.1 - 0.2 (= 0.9) since the complete primary eddy length plus the first length step is equal to 1.1. The probability of occurrence of the primary eddy perturbation at this *t* value however, is determined by the fractional volume dilution *k* which quantifies the departure of the primary eddy from its undiluted average condition and therefore represents the normalized deviation *t*. Therefore the probability density *P* of fractal fluctuations of the primary eddy is given using the computed value of *k* (Eq. 26) as shown in the following equation.

$$P = \tau^{-4k} \qquad (27)$$

The probabilities of occurrence (*P*) of the primary eddy for a complete eddy cycle either in the positive or negative direction are given for progressive growth stages (*t* values) in the following Table 1. The statistical normal probability density distribution corresponding to the normalized deviation *t* values are also given in the Table 1.

| Table 1: Primary eddy growth ||||  |
|---|---|---|---|---|
| Growth step no z | ±t | k | Probability (%) ||
| | | | Model predicted | Statistical normal |
| 2 | .9000 | .8864 | 18.1555 | 18.4060 |
| 3 | .8000 | .7237 | 24.8304 | 21.1855 |
| 4 | .7000 | .6268 | 29.9254 | 24.1964 |
| 5 | .6000 | .5606 | 33.9904 | 27.4253 |
| 6 | .5000 | .5118 | 37.3412 | 30.8538 |
| 7 | .4000 | .4738 | 40.1720 | 34.4578 |
| 8 | .3000 | .4432 | 42.6093 | 38.2089 |
| 9 | .2000 | .4179 | 44.7397 | 42.0740 |
| 10 | .1000 | .3964 | 46.6250 | 46.0172 |
| 11 | 0 | .3780 | 48.3104 | 50.0000 |

The model predicted probability density distribution $P$ along with the corresponding statistical normal distribution with probability values plotted on linear and logarithmic scales respectively on the left and right hand sides are shown in Fig. 2. The model predicted probability distribution $P$ for fractal space-time fluctuations is very close to the statistical normal distribution for normalized deviation $t$ values less than 2 as seen on the left hand side of Fig. 2. The model predicts progressively higher values of probability $P$ for values of $t$ greater than 2 as seen on a logarithmic plot on the right hand side of Fig. 2 and may explain the reported *fat tail* for probability distributions of various physical parameters (Buchanan, 2004).

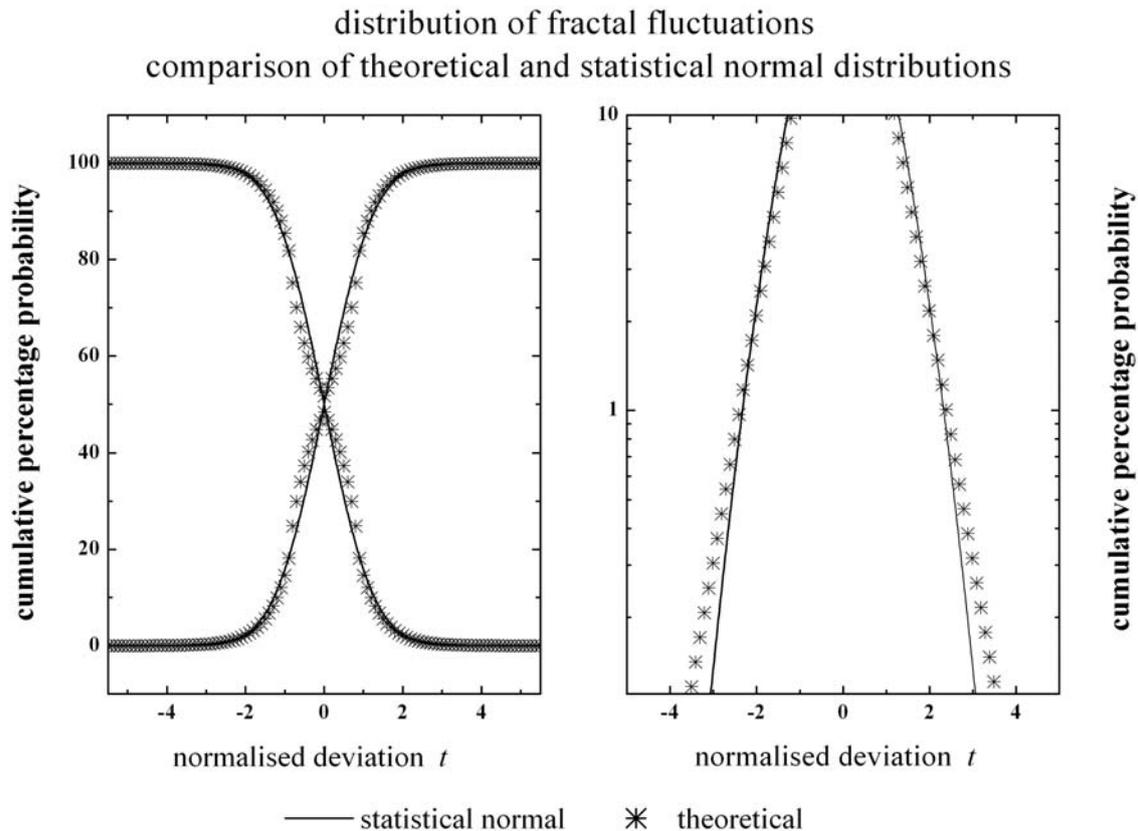

Fig. 2: Probability distribution of fractal fluctuations. Comparison of theoretical with statistical normal distribution.

The probability distribution and the power (variance) spectrum of fractal fluctuations follow the same inverse power law $P = \tau^{-4t}$ where $P$ is the probability density and $t$ is the normalized deviation equal to $(av - x)/sd$ where $av$ and $sd$ are respectively the average and standard deviation of the fractal data series. The probability density $P$ also represents the normalized variance and corresponding normalized deviation $t$ equal to $[\log (L)/\log (T_{50}) - 1]$ where $L$ is the wavelength (period) and $T_{50}$ the wavelength (period) up to which the cumulative percentage contribution to total variance is equal to 50. The corresponding phase spectrum also follows the same inverse power law since eddy circulations are associated with phase angle equal to $r/R$ in Eq. (1) and represent the variance spectrum. Long-range space-time correlations are inherent to inverse power-law distributions. The model predicted probability density $P$ is very close to the statistical normal distribution for normalized deviation $t$ values less than 2, i.e. moderate amplitude fluctuations. For larger amplitude fluctuations, i.e., $t > 2$ the model predicted probability density $P$ is progressively larger than the corresponding statistical normal distribution. The applicability of statistical normal distribution for fractal fluctuations was discussed in Section 2.1 above. The above model prediction, namely that the additive amplitudes of eddies when squared (variance) represent probability densities of eddy fluctuations (amplitudes) is exhibited by the sub-atomic dynamics of quantum systems such as the electron or photon. Therefore, fractal fluctuations exhibit quantum-like chaos.

## 3. General Systems Theory and Classical Statistical Physics

Nature has a hierarchical structure, with time, length and energy scales ranging from the submicroscopic to the supergalactic. Surprisingly it is possible and in many cases essential to discuss these levels independently—quarks are irrelevant for understanding protein folding and atoms are a distraction when studying ocean currents. Nevertheless, it is a central lesson of science, very successful in the past three hundred years, that there are no new fundamental laws, only new phenomena, as one goes up the hierarchy. Thus, arrows of explanations between different levels always point from smaller to larger scales, although the origin of higher level phenomena in the more fundamental lower level laws is often very far from transparent. Statistical Mechanics (SM) provides a framework for describing how well-defined higher level patterns or behavior may result from the non-directed activity of a multitude of interacting lower level individual entities. The subject was developed for, and has had its greatest success so far in, relating mesoscopic and macroscopic thermal phenomena to the microscopic world of atoms and molecules. Statistical mechanics explains how macroscopic phenomena originate in the cooperative behavior of these microscopic particles (Lebowitz, 1999).

The general systems theory visualizes the self-similar fractal fluctuations to result from a hierarchy of eddies, the larger scale being the space-time average of enclosed smaller scale eddies (Eq. 1) assuming constant values for the characteristic length scale $r$ and circulation speed $w_*$ throughout the large eddy space-time domain. The collective behavior of the ordered hierarchical eddy ensembles is manifested as the apparently irregular fractal fluctuations with long-range space-time correlations generic to dynamical systems. The concept that aggregate averaged eddy ensemble properties represent the eddy continuum belongs to 19$^{th}$ century classical statistical physics where the study of the properties of a system is reduced to a determination of average values of the physical quantities that characterize the state of the system as a whole (Yavorsky and Detlaf, 1975) such as gases, e.g., the gaseous envelope of the earth, the atmosphere.

In classical statistical physics *kinetic theory of ideal gases* is a study of systems consisting of a great number of molecules, which are considered as bodies having a small size and mass[33] (Kikoin and Kikoin, 1978). Classical statistical methods of investigation[33-40] (Kikoin and Kikoin, 1978; Dennery, 1972; Yavorsky and Detlaf, 1975; Rosser, 1985; Guenault, 1988; Gupta, 1990; Dorlas, 1999; Chandrasekhar, 2000) are employed to estimate average values of quantities characterizing aggregate molecular motion such as mean velocity, mean energy etc., which determine the macro-scale characteristics of gases. The mean properties of ideal gases are calculated with the following assumptions. (1) The intra-molecular forces are completely absent instead of being small. (2) The dimensions of molecules are ignored, and considered as material points. (3) The above assumptions imply the molecules are completely free, move rectilinearly and uniformly as if no forces act on them. (4) The ceaseless chaotic movements of individual molecules obey Newton's laws of motion.

The Austrian physicist Ludwig Boltzmann suggested that knowing the probabilities for the particles to be in any of their various possible configurations would enable to work out the overall properties of the system. Going one step further, he also made a bold and insightful guess about these probabilities - that any of the many conceivable configurations for the particles would be equally probable. Boltzmann's idea works, and has enabled physicists to make mathematical models of thousands of real materials, from simple crystals to superconductors. It reflects the fact that many quantities in nature - such as the velocities of molecules in a gas - follow "normal" statistics. That is, they are closely grouped around the average value, with a "bell curve" distribution. Boltzmann's guess about equal probabilities only works for systems that have settled down to equilibrium, enjoying, for example, the same temperature throughout. The theory fails in any system where destabilizing external sources of energy are at work, such as the haphazard motion of turbulent fluids or the fluctuating energies of cosmic rays. These systems don't follow normal statistics, but another pattern instead (Buchanan, 2005).

Cohen (2005) discusses Boltzmann's equation as follows. In 1872 when Boltzmann derived in his paper: *Further studies on thermal equilibrium between gas molecules* (Boltzmann, 1872), what we now call the Boltzmann equation, he used, following Clausius and Maxwell, the assumption of 'molecular chaos', and he does not seem to have realized the statistical, i.e., probabilistic nature of this assumption, i.e., of the assumption of the independence of the velocities of two molecules which are going to collide. He used both a dynamical and a statistical method. However, Einstein strongly disagreed with Boltzmann's statistical method, arguing that a statistical description of a system should be based on the dynamics of the system. This opened the way, especially for complex systems, for other than Boltzmann statistics. It seems that perhaps a combination of dynamics and statistics is necessary to describe systems with complicated dynamics (Cohen, 2005). Sornette (2007) discusses the ubiquity of observed power law distributions in complex systems as follows. The extension of Boltzmann's distribution to out-of-equilibrium systems is the subject of intense scrutiny. In the quest to characterize complex systems, two distributions have played a leading role: the normal (or Gaussian) distribution and the power law distribution. Power laws obey the symmetry of scale invariance. Power law distributions and more generally regularly varying distributions remain robust functional forms under a large number of operations, such as linear combinations, products, minima, maxima, order statistics, powers, which may also explain their ubiquity and attractiveness. Research on the origins of power law relations, and efforts to observe and validate them in the real world, is extremely active in many fields of modern science, including physics, geophysics, biology, medical sciences, computer science, linguistics, sociology, economics and more. Power law distributions incarnate the notion that extreme events are not exceptional. Instead, extreme events should be considered as rather frequent and part of the same organization as the other events (Sornette, 2007).

In the following it is shown that the general systems theory concepts are equivalent to Boltzmann's postulates and the *Boltzmann distribution* with the inverse power law expressed as a function of the golden mean is the universal probability distribution function for the observed fractal fluctuations which corresponds closely to statistical normal distribution for moderate amplitude fluctuations and exhibit a fat long tail for hazardous extreme events in dynamical systems.

For any system large or small in thermal equilibrium at temperature $T$, the probability $P$ of being in a particular state at energy $E$ is proportional to $e^{-\frac{E}{k_B T}}$ where $k_B$ is the

*Boltzmann's constant*. This is called the *Boltzmann distribution* for molecular energies and may be written as

$$P \propto e^{-\frac{E}{k_B T}} \qquad (28)$$

The basic assumption that the space-time average of a uniform distribution of primary small scale eddies results in the formation of large eddies is analogous to Boltzmann's concept of equal probabilities for the microscopic components of the system (Buchanan, 2005). The physical concepts of the general systems theory (Section 2) enables to derive (Selvam, 2002) *Boltzmann distribution* as shown in the following.

The r.m.s circulation speed $W$ of the large eddy follows a logarithmic relationship with respect to the length scale ratio $z$ equal to $R/r$ (Eq. 3) as given below

$$W = \frac{w_*}{k} \log z$$

In the above equation the variable $k$ represents for each step of eddy growth, the fractional volume dilution of large eddy by turbulent eddy fluctuations carried on the large eddy envelope (Selvam, 1990) and is given as (Eq. 25)

$$k = \frac{w_* r}{WR}$$

Substituting for $k$ in Eq. (3) we have

$$W = w_* \frac{WR}{w_* r} \log z = \frac{WR}{r} \log z$$

and  (29)

$$\frac{r}{R} = \log z$$

The ratio $r/R$ represents the fractional probability $P$ of occurrence of small-scale fluctuations ($r$) in the large eddy ($R$) environment. Since the scale ratio $z$ is equal to $R/r$ Eq. 29 may be written in terms of the probability $P$ as follows.

$$\frac{r}{R} = \log z = \log\left(\frac{R}{r}\right) = \log\left(\frac{1}{(r/R)}\right)$$

$$P = \log\left(\frac{1}{P}\right) = -\log P \qquad (30)$$

The maximum entropy principle concept of classical statistical physics is applied to determine the fidelity of the inverse power law probability distribution $P$ (Eq. 24) for exact quantification of the observed space-time fractal fluctuations of dynamical systems ranging from the microscopic dynamics of quantum systems to macro-scale real world systems. Kaniadakis (2009) states that the correctness of an analytic expression for a given power-law tailed distribution, used to describe a statistical system, is strongly related to the validity of the generating mechanism. In this sense the maximum entropy principle, the cornerstone of statistical physics, is a valid and powerful tool to explore new roots in searching for generalized statistical theories (Kaniadakis, 2009). The concept of entropy is fundamental in the foundation of statistical physics. It first appeared in thermodynamics through the second law of thermodynamics. In statistical mechanics, we are interested in the disorder in the distribution of the system over the permissible microstates. The measure of disorder first provided by Boltzmann principle (known as Boltzmann entropy) is given by $S = K_B \ln M$, where $K_B$ is the thermodynamic unit of measurement of entropy and is known as Boltzmann constant. $K_B = 1.33 \times 10^{-16}$ erg/°C. $M$, called thermodynamic probability or statistical weight, is the total number of microscopic complexions compatible with the macroscopic state of the system and corresponds to the "degree of disorder" or 'missing information' (Chakrabarti and De, 2000). For a probability distribution among a discrete set of states the generalized entropy for a system out of equilibrium is given as (Salingaros and West, 1999; Chakrabarti and De, 2000; Beck, 2009; Sethna, 2009)

$$S = -\sum_{j=1}^{n} P_j \ln P_j \tag{31}$$

In Eq. (31) $P_j$ is the probability for the $j^{th}$ stage of eddy growth in the present study and the entropy $S$ represents the 'missing information' regarding the probabilities. Maximum entropy $S$ signifies minimum preferred states associated with scale-free probabilities.

The validity of the probability distribution $P$ (Eq.24) is now checked by applying the concept of maximum entropy principle (Kaniadakis, 2009). Substituting for log $P_j$ (Eq.30) and for the probability $P_j$ in terms of the golden mean $\tau$ derived earlier (Eq. 24) the entropy $S$ is expressed as

$$S = -\sum_{j=1}^{n} P_j \log P_j = \sum_{j=1}^{n} P_j^2 = \sum_{j=1}^{n} \left(\tau^{-4n}\right)^2$$
$$S = \sum_{j=1}^{n} \tau^{-8n} \approx 1 \text{ for large } n \tag{32}$$

In Eq. (32) $S$ is equal to the square of the cumulative probability density distribution and it increases with increase in $n$, i.e., the progressive growth of the eddy continuum and approaches 1 for large $n$. According to the second law of thermodynamics, increase in entropy signifies approach of dynamic equilibrium conditions with scale-free characteristic of fractal fluctuations and hence the probability distribution $P$ (Eq. 24) is the correct analytic expression quantifying the eddy growth processes visualized in the general systems theory.

In the following it is shown that the eddy continuum energy distribution $P$ (Eq. 24) is the same as the *Boltzmann distribution* for molecular energies. From Eq. (29)

$$z = \frac{R}{r} = e^{\frac{r}{R}}$$
or
$$\frac{r}{R} = e^{-\frac{r}{R}} \tag{33}$$

The ratio $r/R$ represents the fractional probability $P$ (Eq. 24) of occurrence of small-scale fluctuations ($r$) in the large eddy ($R$) environment. Considering two large eddies of radii $R_1$ and $R_2$ ($R_2$ greater than $R_1$) and corresponding r.m.s circulation speeds $W_1$ and $W_2$ which grow from the same primary small-scale eddy of radius $r$ and r.m.s circulation speed $w_*$ we have from Eq. (1)

$$\frac{R_1}{R_2} = \frac{W_2^2}{W_1^2}$$

From Eq. (33)

$$\frac{R_1}{R_2} = e^{-\frac{R_1}{R_2}} = e^{-\frac{W_2^2}{W_1^2}} \qquad (34)$$

The square of r.m.s circulation speed $W^2$ represents eddy kinetic energy. Following classical physical concepts (Kikoin and Kikoin, 1978) the primary (small-scale) eddy energy may be written in terms of the eddy environment temperature $T$ and the *Boltzmann's constant* $k_B$ as

$$W_1^2 \propto k_B T \qquad (35)$$

Representing the larger scale eddy energy as $E$

$$W_2^2 \propto E \qquad (36)$$

The length scale ratio $R_1/R_2$ therefore represents fractional probability $P$ (Eq. 24) of occurrence of large eddy energy $E$ in the environment of the primary small-scale eddy energy $k_B T$ (Eq. 35). The expression for $P$ is obtained from Eq. (34) as

$$P \propto e^{-\frac{E}{k_B T}} \qquad (37)$$

The above is the same as the *Boltzmann's equation* (Eq. 28).

The derivation of *Boltzmann's equation* from general systems theory concepts visualises the eddy energy distribution as follows: (1) The primary small-scale eddy represents the molecules whose eddy kinetic energy is equal to $k_B T$ as in classical physics. (2) The energy pumping from the primary small-scale eddy generates growth of progressive larger eddies (Selvam, 1990). The r.m.s circulation speeds $W$ of larger eddies are smaller than that of the primary small-scale eddy (Eq. 1). (3) The space-time *fractal* fluctuations of molecules (atoms) in an ideal gas may be visualized to result from an eddy continuum with the eddy energy $E$ per unit volume relative to primary molecular kinetic energy ($k_B T$) decreasing progressively with increase in eddy size.

The eddy energy probability distribution ($P$) of fractal space-time fluctuations also represents the *Boltzmann distribution* for each stage of hierarchical eddy growth and is given by Eq. (24) derived earlier, namely

$$P = \tau^{-4t}$$

The general systems theory concepts are applicable to all space-time scales ranging from microscopic scale quantum systems to macroscale real world systems such as atmospheric flows.

## 4. Data sets used for the study

Monthly total rainfalls for the period 1900 to 2008 for all available Indian and USA stations were obtained from Global Historical Climatology Network (GHCN) of the National Oceanic and Atmospheric Administration's National Climate Data Center, Version 2 Precipitation Version 2 data sets, raw data (v2.prcp) (http://www.ncdc.noaa.gov/oa/climate/ghcn-monthly/index.php). The data are monthly total precipitation recorded at the station in tenths of mm. The annual total rainfall was computed for the years where all the 12 months rainfall data is available. A total of 504 Indian and 764 USA stations where continuous rainfall data for a minimum of 50-years was available were considered for the study. The mean, standard deviation and the number of years of available rainfall data for the Indian and USA stations are shown in Fig. 3.

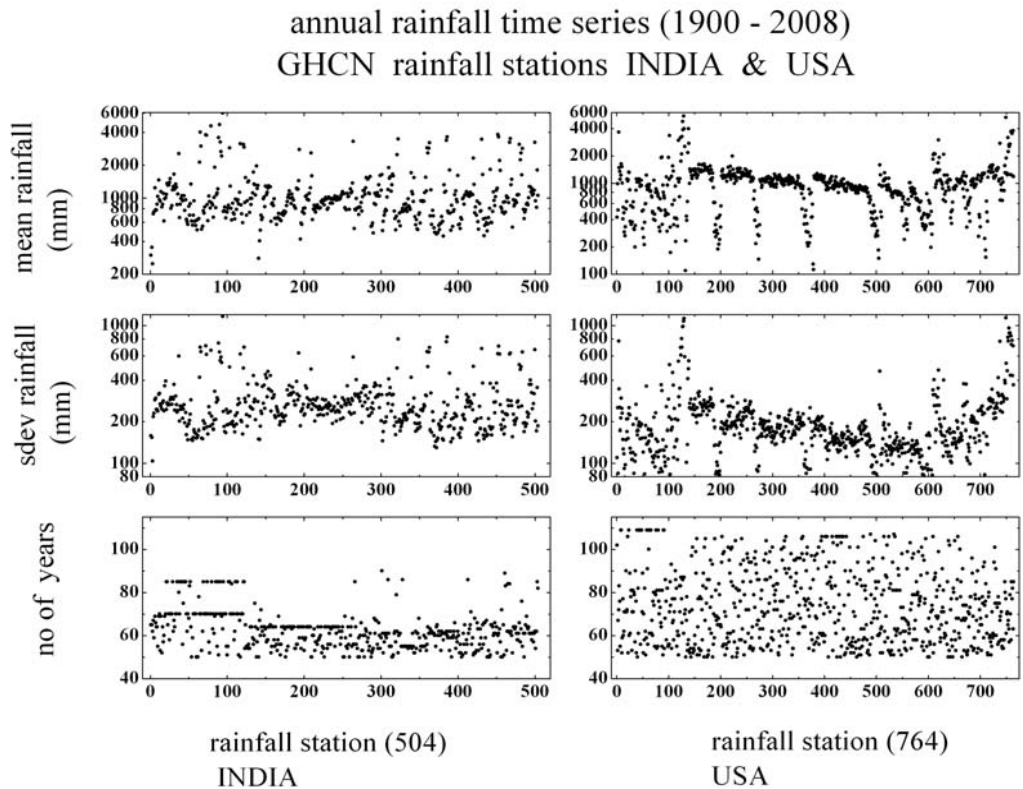

Fig. 3. The mean, standard deviation and the number of years of available rainfall data for the Indian and USA region stations.

## 5. Analyses and Results

### 5.1 Frequency distribution

For each station rainfall time series $x(i)$, $i=1, n$ where $x(i)$ is the annual total rainfall for the year $i$ and $n$ the total number of years, the mean ($av$) and the standard deviation ($sd$) were computed. The rainfall amounts for the $n$ years were then arranged in ascending order of magnitude ranging from $x_{min}$ to $x_{max}$, the respective minimum and maximum rainfall amounts. The ascending order rainfall amounts sequence is then expressed in terms of $m$ values of the normalized deviation $t(i)$ equal to $(x(i)-av)/sd$ ranging from the smallest value $t_{min}$ to the largest value $t_{max}$. The value of time series length $n$ is equal to the $t(i)$ sequence length $m$ for most of the stations since the frequency $f(i)$ of occurrence of each $t(i)$ value is equal to one in most cases except for a few stations where rainfall amounts for two or more years are the same. The cumulative percentage frequencies of occurrence $cmax(i)$ and $cmin(i)$ corresponding to the *normalized deviation t* values were then computed starting respectively from the maximum ($t_{max}$) and minimum ($t_{min}$) $t$ values.

$$cmax(i) = \frac{\sum_{m}^{i} f(i)}{\sum_{1}^{m} f(i)} \times 100 = \frac{\sum_{m}^{i} f(i)}{n} \times 100$$

Similarly

$$cmin(i) = \frac{\sum_{1}^{i} f(i)}{n} \times 100$$

The average cumulative percentage frequencies of occurrence $cmax(i)$ and $cmin(i)$ with respect to the corresponding $t$ values for the 504 Indian and 764 USA rainfall stations are shown in Fig. 4.

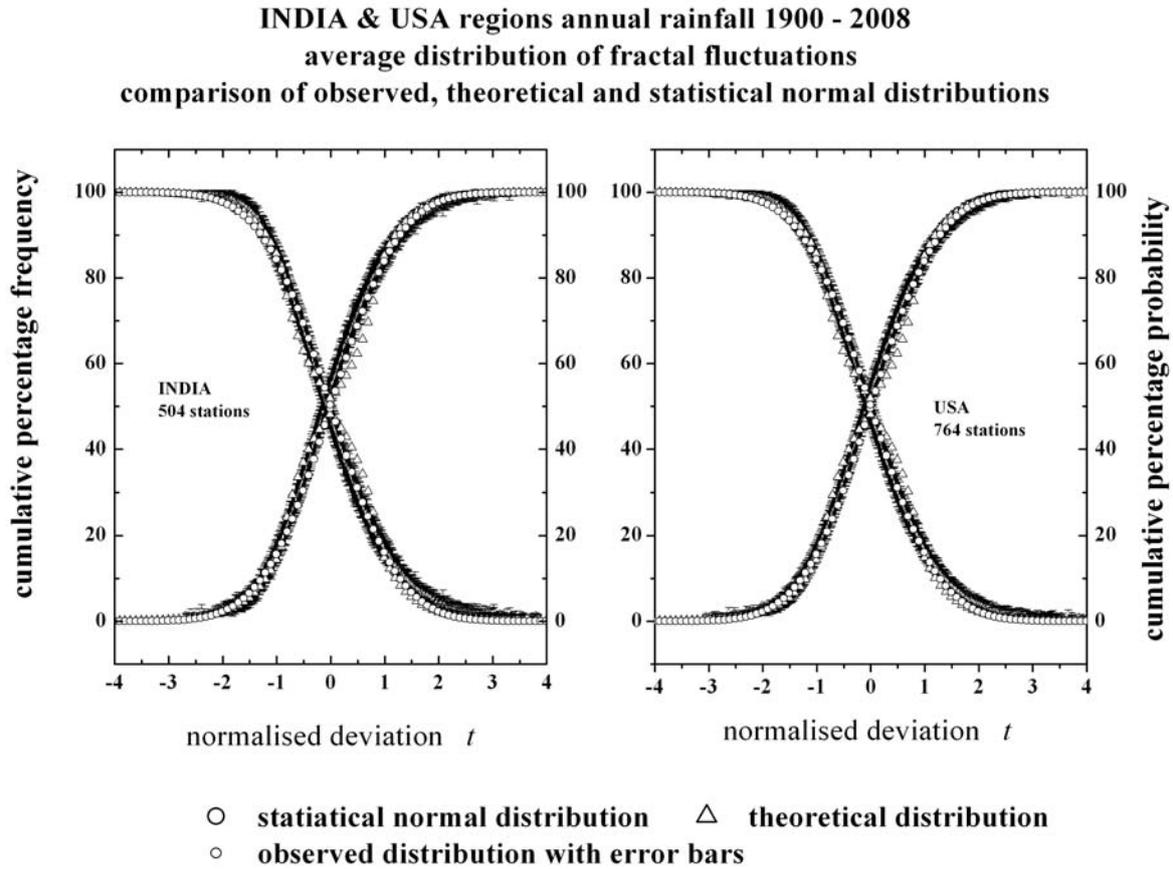

Fig. 4. Average cumulative percentage probability distributions (negative to positive tail and vice-versa) for Indian and USA region stations annual rainfall time series. The error bars indicate one standard deviation on either side of the mean. The model predicted theoretical distribution and the statistical normal distribution are also shown in the figure.

The average cumulative percentage probability values $cmax(i)$ and $cmin(i)$ for Indian and USA region stations are plotted with respect to corresponding *normalized deviation t* values on logarithmic scale for the probability axis in the tail regions, i.e. $t$ values greater than 2 in Fig. 5. The positive extremes $t = 2$ to $4$ and the negative extremes $t = -2$ to $-4$ are shown respectively on the left and right side of Fig. 5. The standard deviation of each mean $cmax(i)$ and $cmin(i)$ value is shown as a vertical error bar on either side of the mean in Fig. 5. The figure also contains the statistical normal distribution and the computed theoretical distribution (Eq. 24) for comparison.

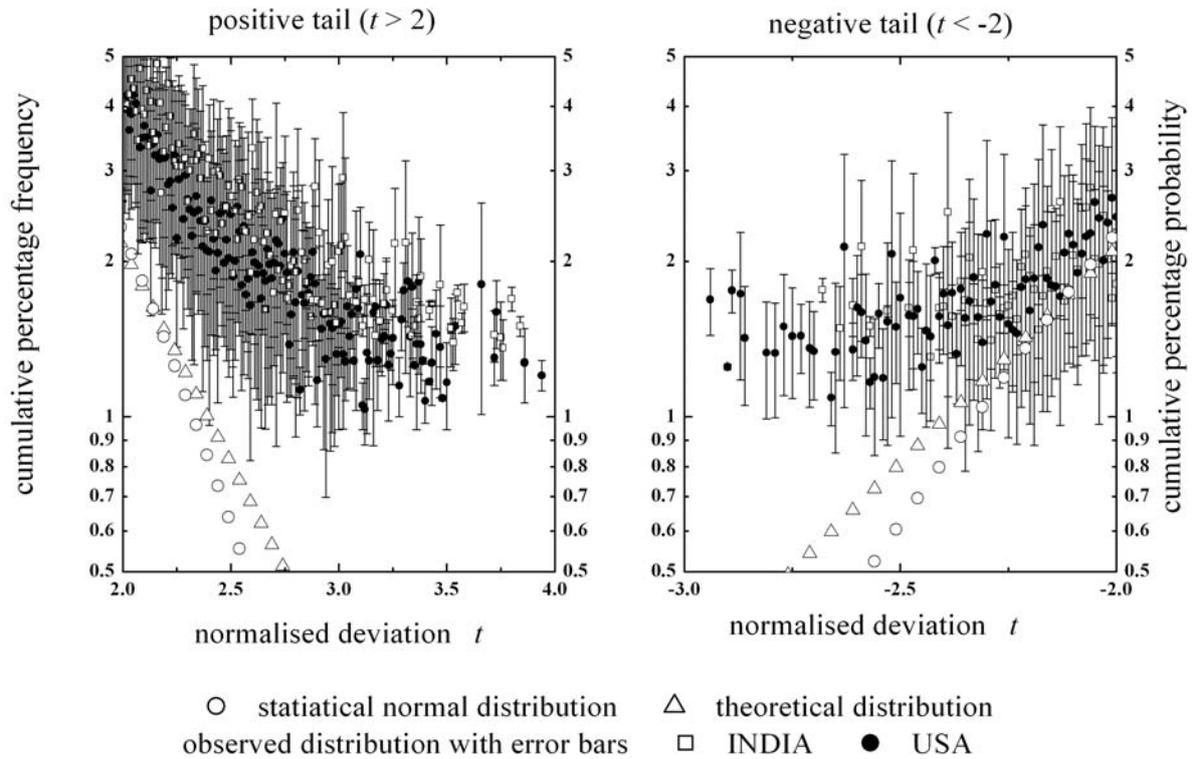

Fig. 5. Average cumulative percentage probability values on logarithmic scale for the probability axis in the tail regions (extreme events). Positive ($t > 2$) and negative ($t > -2$) tail regions are shown on the left and right side respectively in the above figure. The model predicted probability distribution and the statistical normal distribution are also shown in the figure.

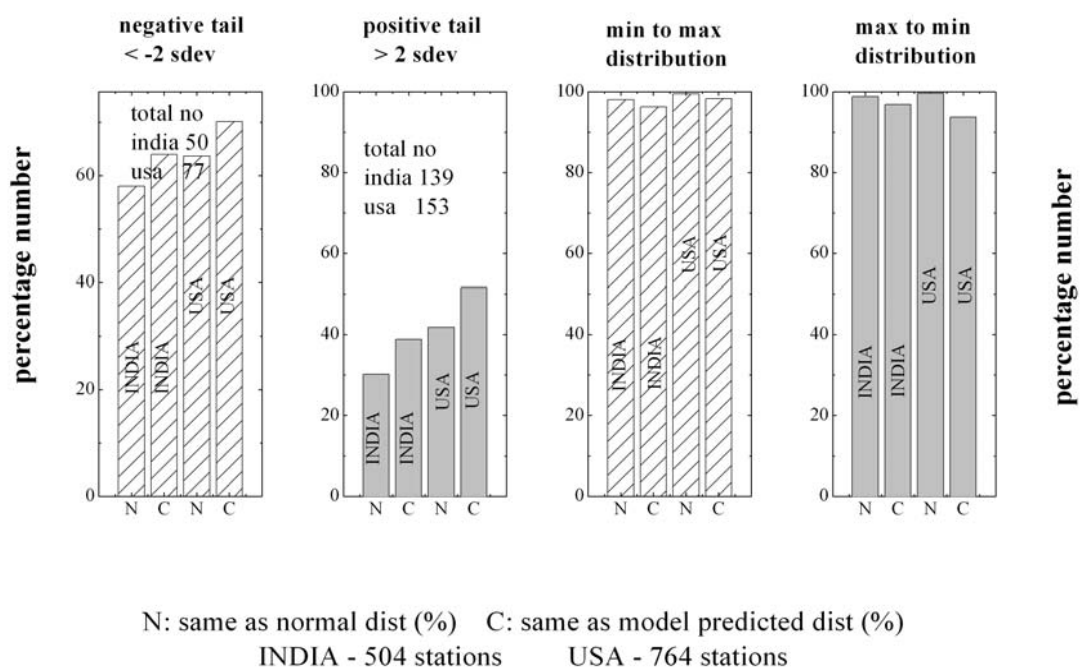

Fig. 6. The first two histograms give the percentage number of extreme values with cumulative percentage probability of occurrence same as (i) statistical normal (N) and (ii) same as theoretical distribution (C) for negative tail ($t < -2$) and positive tail ($t > 2$) regions. The last two histograms give the percentage number of the cumulative probability distributions computed (i) starting from minimum $cmin(i)$ and (ii) starting from maximum $cmax(i)$ same as statistical normal (N) and same as theoretical distribution (C)

The Fig. 5 show clearly the appreciable positive departure of observed probability densities from the statistical normal distribution for extreme values at normalized deviation $t$ values more than 2. The observed extreme values corresponding to $t$ values greater than 2 for $cmax(i)$ and $cmin(i)$ distributions were compared for 'goodness of fit' with computed theoretical distribution and statistical normal distribution as follows. For $cmax(i)$ and $cmin(i)$ values, where standard deviation is available (number of observed values more than one), if the observed distribution included the computed theoretical (statistical normal) distribution within twice the standard deviation on either side of the mean then it was assumed to be the same as the computed theoretical (statistical normal) distribution at 5% level of significance within measurement errors. The number of observed distribution values which included the computed theoretical values and/or the statistical normal distribution values within twice the standard deviation on either side of the mean was determined. The total and percentage numbers of observed extreme values same as computed theoretical and statistical normal distributions are given in Fig. 6 for positive and negative tail regions (normalized deviation $t$ greater than 2) for India and USA. The percentage number of observed extreme values with same probability as model predicted (computed) is more than the percentage number of extreme values with same probability as statistical normal distribution for India and USA (first and second histograms in Fig. 6). More than 90% of the observed cumulative probability distributions computed starting from either end (minimum or maximum) are the same as the model predicted theoretical and also the statistical normal distribution (third and fourth histogram in Fig. 6) and such a result is consistent since the model predicts significantly larger probability values only in the tail regions with normalized deviation *t* values greater than two.

## 5.2 Continuous periodogram power spectral analyses

The power spectra of frequency distribution of monthly mean data sets were computed accurately by an elementary, but very powerful method of analysis developed by Jenkinson (1977) which provides a quasi-continuous form of the classical periodogram allowing systematic allocation of the total variance and degrees of freedom of the data series to logarithmically spaced elements of the frequency range (0.5, 0). The cumulative percentage contribution to total variance was computed starting from the high frequency side of the spectrum. The corresponding phase spectrum was computed as equal to the percentage contribution to total rotation. The average power (variance) and phase spectra are plotted for Indian and USA regions as cumulative percentage contribution to total variance versus the normalized standard deviation *t* equal to $(\log L/\log T_{50}) - 1$ where *L* is the period in years and $T_{50}$ is the period up to which the cumulative percentage contribution to total variance is equal to 50 (Eq. 7). The statistical *Chi-Square* test (Spiegel, 1961) was applied to determine the 'goodness of fit' of individual variance and phase spectrum with each other and also with statistical normal distribution and model predicted variance spectrum (Eqs. 21 and 24). The variance and corresponding phase spectra covered the range of normalized deviation *t* values from 3 to -1 and the average spectra shown in Fig. 7 for Indian and USA regions are found to follow closely the statistical normal distribution. The percentage numbers of (i) variance (phase) spectra same as normal distribution (ii) variance (phase) spectra same as theoretical distribution and (iii) variance spectra same as corresponding phase spectra are shown in Fig. 8. Variance spectra show closer correspondence with normal distribution and theoretical distribution than phase spectra. Variance spectra same as normal and computed distributions respectively are nearly 100% and greater than 90% (first histogram in Fig. 8), phase spectra same as normal and computed distributions respectively are more than 80% and 70% (second histogram in Fig. 8), variance spectra same as corresponding phase spectra is more than 85% (third histogram in Fig. 8).

The values of period $T_{50}$ up to which the cumulative percentage contribution to total variance is equal to 50 plotted in Fig. 9 for Indian and USA stations are close to the model predicted value $T_{50} = 3.6$ years (Eq. 7).

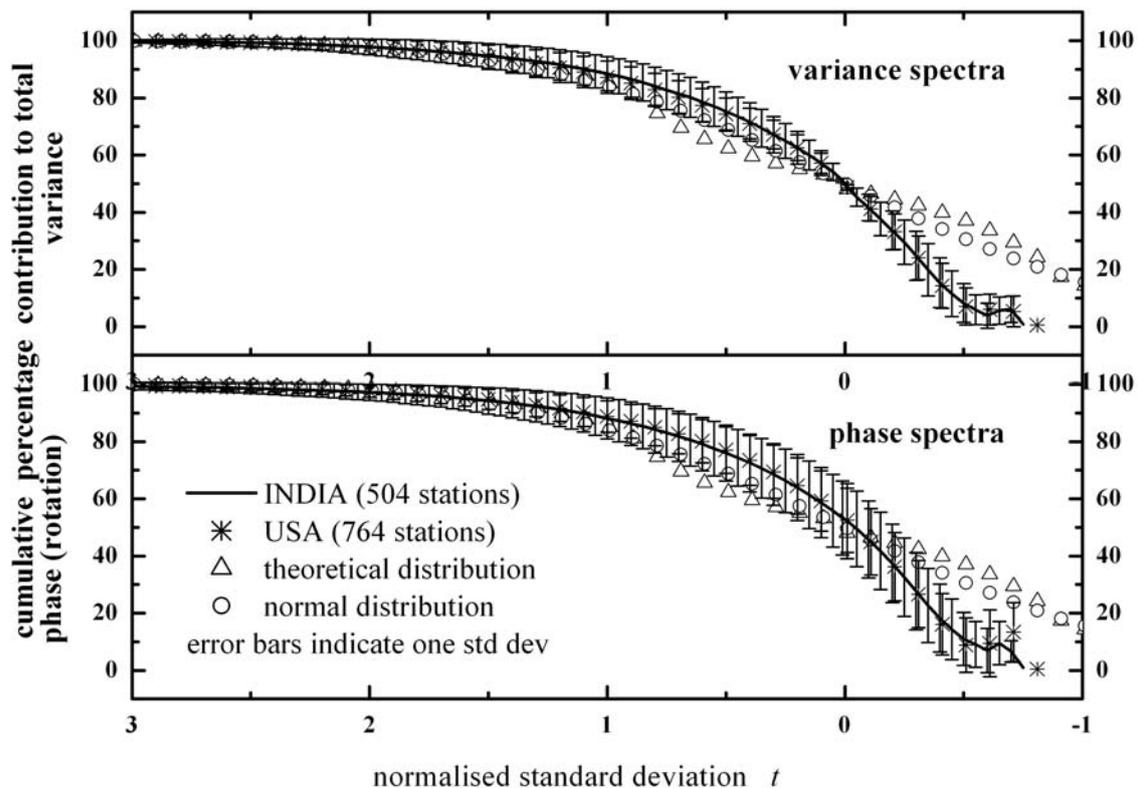

Fig. 7. Average variance and phase spectra for Indian and USA region annual rainfall time series (1900-2008). Error bars indicate one standard deviation on either side of mean. The statistical normal distribution and the theoretical distribution are also shown in the figure.

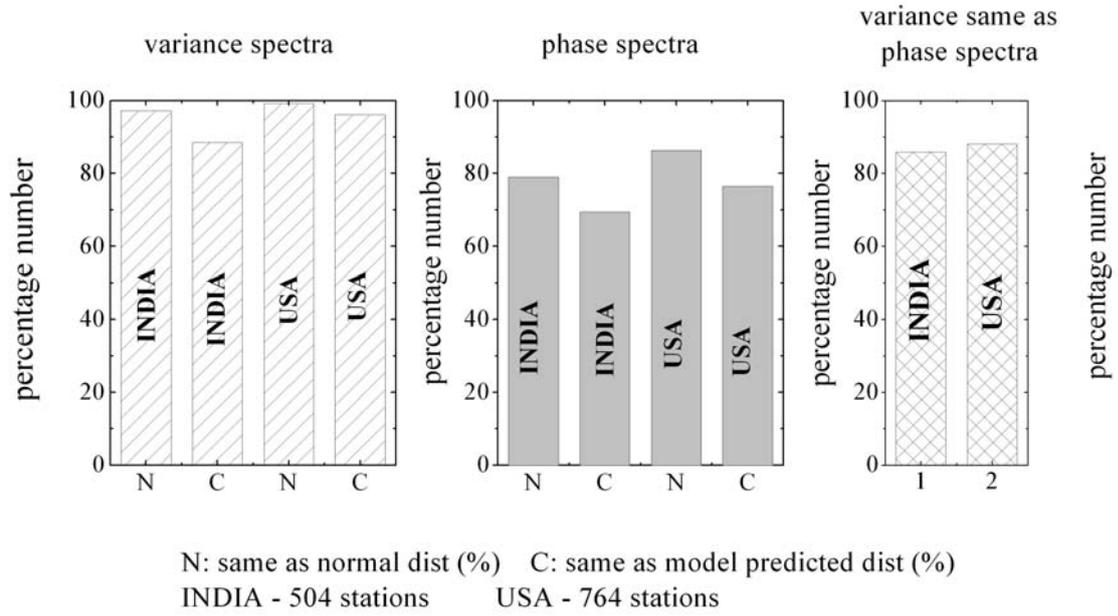

Fig. 8. The percentage numbers of variance and phase spectra same as (i) normal distribution (N) (ii) theoretical distribution (C) and (iii) variance spectra same as phase spectra for India and USA region rainfall stations.

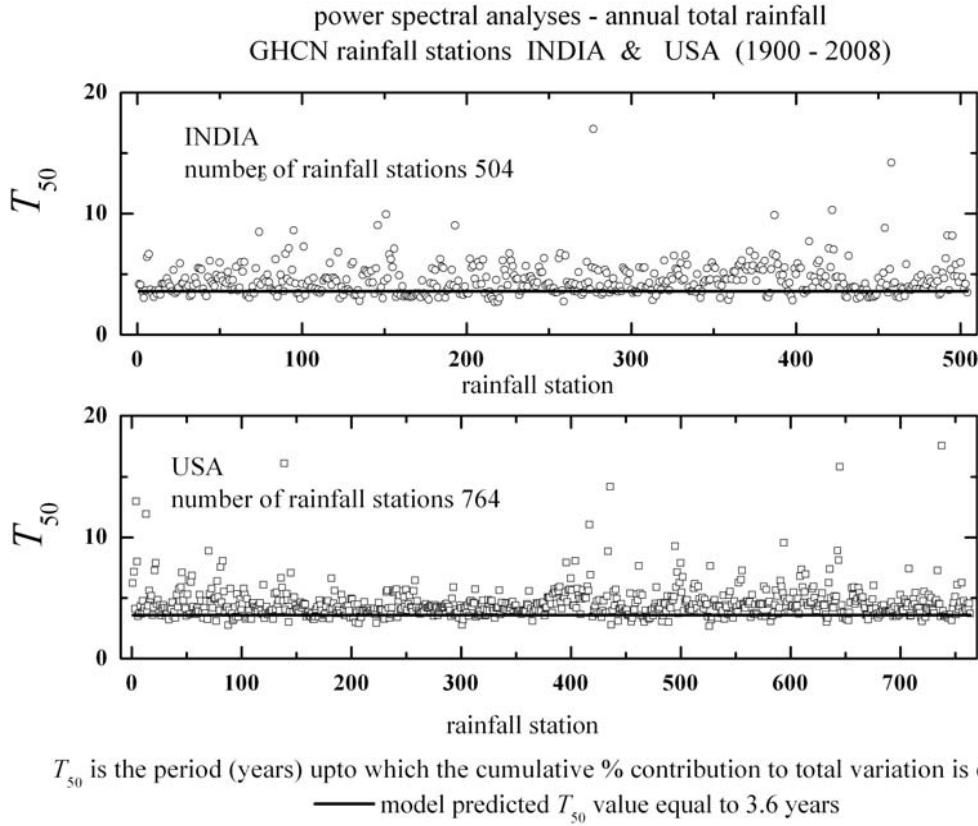

Fig. 9. The period $T_{50}$ up to which the cumulative percentage contribution to total variance is equal to 50 for the Indian and USA rainfall time series. The horizontal line is the model predicted $t_{50}$ equal to 3.6 years for interannual variability of rainfall corresponding to the annual (one year) summer to winter cycle of solar heating of the atmosphere.

## 6. Discussion and conclusions

Dynamical systems in nature exhibit selfsimilar fractal fluctuations for all space-time scales and the corresponding power spectra follow inverse power law form signifying long-range space-time correlations identified as self-organized criticality (Bak, et al., 1988). The physics of self-organized criticality is not yet identified. The Gaussian probability distribution used widely for analysis and description of large data sets is found to significantly underestimate the probabilities of occurrence of extreme events such as stock market crashes, earthquakes, heavy rainfall, etc. Further, the assumptions underlying the normal distribution such as fixed mean and standard deviation, independence of data, are not valid for real world fractal data sets exhibiting a scale-free power law distribution with fat tails. It is important to identify and quantify the fractal distribution characteristics of dynamical systems for predictability studies.

A recently developed general systems theory for fractal space-time fluctuations (Selvam, 1990, 2005, 2007, 2009, 2010; Selvam and Fadnavis, 1998) shows that the larger scale fluctuation can be visualized to emerge from the space-time averaging of enclosed small scale fluctuations, thereby generating a hierarchy of self-similar fluctuations manifested as the observed eddy continuum in power spectral analyses of fractal fluctuations. The concept that aggregate averaged eddy ensemble properties represent the eddy continuum belongs to 19[th] century classical statistical physics where the study of the properties of a system is reduced to a determination of average values of the physical quantities that characterize the state of the system as a whole (Yavorsky and Detlaf, 1975) such as gases, e.g., the gaseous envelope of the earth, the atmosphere. The basic assumption that the space-time average of a uniform distribution of primary small scale eddies results in the formation of large eddies is analogous to Boltzmann's concept of equal probabilities for the microscopic components of the system. The physical concepts of the general systems theory (Section 2) enables to derive (Selvam, 2002, 2010) the universal inverse power law for fractal fluctuations in the form of the *Boltzmann distribution* where the probability distribution function $P$ for fractal fluctuations follow inverse power law form $\tau^{-4t}$ where $\tau$ is the *golden mean*, and $t$, the normalized deviation is equal to ($x$-$av$/$sd$) where $av$ and $sd$ are respectively the average and standard deviation of the distribution. The apparently disordered (irregular) fractal fluctuations self-organize to maintain a dynamical equilibrium state, namely, the universal inverse power law distribution, thereby fulfilling the second law of thermodynamics. The predicted distribution is close to the Gaussian distribution for small-scale fluctuations (normalized deviation $t$ less than 2), but exhibits *fat long tail* for large-scale fluctuations (normalized deviation $t$ more than 2) with higher probability of occurrence than predicted by Gaussian distribution. There is always a non-zero probability of occurrence of very large amplitude, damage causing extreme events.

The model predicts the same probability distribution $P$ for the amplitude as well as the power (variance) spectrum of fractal fluctuations. Such a result that the additive amplitudes of eddies when squared represent probabilities is exhibited by the sub-atomic dynamics of quantum systems such as the electron or photon (Maddox, 1988, 1993; Rae, 1988). Fractal fluctuations therefore exhibit quantum-like chaos. The non-dimensional fine structure constant of the universal inverse power law spectrum ($P$) of fractal fluctuations is equal to about 1/137 in close agreement with that observed for atomic spectra.

Analysis of historic (1900 -2008) data sets of annual precipitation (GHCN V2. prcp) time series for all available stations in India and USA show that the data follow closely, but not exactly the statistical normal and the model predicted distributions in the region of normalized deviations $t$ less than 2 (Fig. 4). For normalized deviations $t$ greater than 2, the data exhibit significantly larger probabilities as compared to the normal distribution and closer to the model predicted probability distribution (Fig 5). A simple $t$ test for 'goodness of fit' of the extreme values (normalized deviation t > 2) of the observed distribution with model predicted (theoretical) and also the statistical normal distribution shows that more number of data points exhibit significant (at 5% level) 'goodness of fit' with the model predicted (theoretical) distribution than with the normal distribution (Fig. 6). The mean power spectra follow closely the statistical normal distribution for India and USA (Fig. 7) region rainfall time series. The power spectra mostly cover the range for *normalized deviation t* less than 2 where the model predicted theoretical distribution is close to the statistical normal distribution. A majority (90% and more) of the power spectra follow closely statistical normal and also to a lesser extent the theoretical distribution (Fig. 8) consistent with model prediction of quantum-like chaos, i.e., variance or square of eddy amplitude represents the probability distribution, a signature of quantum systems. Universal spectrum for inter-annual variability have been reported in the following meteorological parameters: (i) rainfall time series over the Indian Region (Selvam et al., 1992) (ii) rainfall time series over India and the United Kingdom (Selvam et al., 1995) (iii) COADS global air and sea surface temperatures (Selvam and Joshi, 1995) (iv) interannual variability in COADS surface pressure time series (Selvam et al., 1996) (v) interannual variability in some disparate climatic regimes (Selvam and Fadnavis, 1998) (vi) global mean monthly temperature anomalies (Selvam, 2010.

The model predicted and observed universal spectrum for inter-annual variability rules out linear secular trends in annual rainfall over Indian and USA regions. Atmospheric energy input related to global warming results in intensification of fluctuations of all scales and is manifested immediately in high frequency fluctuations such as the increase in frequency of occurrence heavy rainfall/drought. The general systems theory, originally developed for turbulent fluid flows, provides universal quantification of physics underlying fractal fluctuations and is applicable to all dynamical systems in nature independent of its physical, chemical, electrical, or any other intrinsic characteristic.

## Acknowledgement

The author is grateful to Dr.A.S.R.Murty for encouragement.